\documentclass{article}
\usepackage{dcase2019,amsmath,graphicx,url,times,booktabs, tabularx}
\usepackage{cite}
\usepackage{amssymb}% http://ctan.org/pkg/amssymb
\usepackage{pifont}% http://ctan.org/pkg/pifont
\newcommand{\cmark}{\ding{51}}%
\newcommand{\xmark}{\ding{55}}%
\usepackage{xcolor}

\title{Receptive-Field-Regularized CNN Variants for \\Acoustic Scene Classification }

\name{Khaled Koutini$^1$, Hamid Eghbal-zadeh$^{1,2}$, Gerhard Widmer$^{1,2}$}
\address{
$^1$Institute of Computational Perception (CP-JKU) \& $^2$LIT Artificial Intelligence Lab,\\
Johannes Kepler University Linz, Austria\\
}

\begin{document}
 
\ninept
\maketitle
\nocite{Koutini2019Receptive}
\begin{sloppy}

\begin{abstract}
Acoustic scene classification and related tasks have been dominated by Convolutional Neural Networks (CNNs)~\cite{eghbal-zadehCPJKUSubmissionsDCASE20162016,hersheyCNNArchitecturesLargescale2017,lehnerClassifyingShortAcoustic2017,DorferDCASE2018task1,SakashitaDCASE2018task1,DorferDCASE2018task2,IqbalDCASE2018task2,LeeDCASE2017task4,KoutiniDCASE2018task4}. Top-performing CNNs use mainly audio spectograms as input and borrow their architectural design primarily from computer vision. A recent study~\cite{Koutini2019Receptive} has shown that restricting the receptive field (RF) of CNNs in appropriate ways is crucial for their performance, robustness and generalization in audio tasks. One side effect of restricting the RF of CNNs is that more frequency information is lost. In this paper, we perform a systematic investigation of different RF configuration for various CNN architectures
% , following the guidelines of ~\cite{Koutini2019Receptive},
on the DCASE 2019 Task 1.A dataset.
%~\cite{Mesaros2018_DCASE};
Second, we introduce Frequency Aware CNNs to compensate for the lack of frequency information caused by the restricted RF, and experimentally determine if and in what RF ranges they yield additional improvement.
The result of these investigations are several
% This approach enabled us to surcharge ?? sure? strange word ... %stretch or maximize
% he performance of an architecture. Leading to
well-performing submissions to different tasks in the DCASE 2019 Challenge. % ~\cite{dcase2019web}.

% \hamid{we need to rewrite the abstract. this is rather written for the challenge.}\hamid{are we gonna report task1b and task2 results as well?}
% \hamid{possible title: Adjusting neural network architectures for audio classification and tagging using receptive field regularization and frequency aware convolutional layers}
% In this paper, we detail the \emph{CP-JKU} submissions to the DCASE-2019 challenge Task 1 (acoustic scene classification). % and Task 2 (audio tagging with noisy labels and minimal supervision). 
% In all of our submissions, we use fully convolutional deep neural networks architectures that are regularized with Receptive Field (RF) adjustments.
% We adjust the RF of variants of Resnet and Densenet architectures to best fit the various audio processing tasks that use the spectrogram features as input.
% Additionally, we propose a new CNN layer called \emph{Frequency-Aware Convolution}
% and new noise compensation techniques such as \emph{Adaptive Weighting for Learning from Noisy Labels} \gerhard{[this will not be reported here, will it?]} to cope with the complexities of each task. 
% We prepared \emph{all} of our submissions \emph{without the use of any external data}.
% Our focus in this year's submissions is to provide the best-performing \emph{single-model} submission, using our proposed approaches. 
\end{abstract}


\nocite{Koutini2019Receptive}

\section{Introduction}
\label{sec:intro}
%!TEX root = ../main.tex
Convolutional Neural Networks (CNNs) have shown great promise as end-to-end classifiers in many tasks such as image classification~\cite{heDeepResidualLearning2016,huangDenselyConnectedConvolutional2017} and acoustic scene classification~\cite{DorferDCASE2018task1,Koutinitrrfcnns2019}.
Although every year new architectures are proposed that achieve better image recognition performance, we showed in a recent study~\cite{Koutini2019Receptive} that these performance gains do not seem to translate to the audio domain.
As a solution, we proposed regularizing the \emph{receptive field (RF)} of such CNN architectures in specific ways.
The method was applied to several state-of-the-art image recognition architectures, and the resulting models were shown to then achieve state-of-the-art performance in audio classification tasks~\cite{Koutini2019Receptive}.

Although CNNs can learn their own features and build internal representations from data, the details of how they actually function is crucial to their success in a specific task.
In the image recognition domain, a recent study~\cite{brendel2018approximating} shed light on the decision making procedure of CNNs and showed that using occurrences of small local image features without taking into account their spatial ordering, CNNs can still achieve state-of-the-art results.
However, while spatial ordering and local neighboring relations might not be crucial for object recognition in images, this is not the case in audio representations such as spectrograms. %\hamid{Khaled: the next part is related to Sepp's question. Correct it when needed.}
% In audio spectrograms, %\hamid{due to the commonly used non-linear scale, the filters applied, etc,} 
A local pattern in lower frequencies does not represent the same acoustic event as the same pattern appearing in higher frequencies.
Since CNNs with limited receptive fields\footnote{as shown in~\cite{Koutini2019Receptive}, large RFs result in overfitting in audio classification.}
are only capable of capturing local features and unlike models such as capsule networks~\cite{sabour2017dynamic}, they are unable to find spatial relations between these local patterns.
As convolution is equivariant, each filter is applied to the input to generate an output activation, but the output of the network does not know where exactly each filter is.
This means that if a specific pattern appears in lower frequencies, and a very similar pattern appears in higher frequencies, later convolutional layers cannot distinguish between the two, and this can result in vulnerabilities in such cases.

In~\cite{liu2018intriguing}, Liu et al.~analyzed a generic inability of CNNs to map a pixel in a 2D space, to its exact cartesian coordinate. They address this problem by adding an additional channel to the convolutional layers that contains only the pixel coordinates.
Inspired by this solution, we propose a new convolutional layer for audio processing -- the \emph{Frequency-aware Convolutional Layer} -- to cope with the aforementioned problems in CNNs.
We use an additional channel in the convolutional layer that only contains the frequency information which connects each filter to the frequency bin it is applied to.

In this paper, we extend our previous work~\cite{Koutini2019Receptive} by modifying the receptive field (RF) of various new architectures such as Resnet~\cite{heDeepResidualLearning2016}, PreAct ResNet~\cite{He2016preact,preactbn2018}, Shake-shake~\cite{gastaldi2017shake,preactbn2018}, Densenet~\cite{huangDenselyConnectedConvolutional2017}, and our new frequency-aware \emph{FAResNet} according to the guidelines provided in~\cite{Koutini2019Receptive}, 
aiming at pushing the performance of these models on acoustic scene classification tasks. % We achieve this by fine-tuning the RF of CNN architectures over the input spectrograms, and analysing the performance gains. 
%that can achieve good performance with a single model and compare these variants \gerhard{[what?]} with their \gerhard{[whose?]} frequency-aware counterpart
%\gerhard{[one or several?]}.
% the next sentence is repetition
%In~\cite{Koutini2019Receptive}, we analyzed the effect of RF tuning on the performance of various CNN architectures for ASC.
%We extend our previous work~\cite{Koutini2019Receptive} by modifying the RF of various new architectures such as \hamid{Khaled, you can add all the architectures that you used, in here; and cite their original paper.} Resnet~\cite{heDeepResidualLearning2016}, Shake-shake~\cite{gastaldi2017shake}, and Densenet~\cite{huangDenselyConnectedConvolutional2017} according to the guidelines provided in~\cite{Koutini2019Receptive}.
%\hamid{khaled, the rest needs an update.}
Systematic experiments permit us to determine optimal RF ranges for various architectures on the DCASE 2019 datasets.
We show that configuring CNNs to have a receptive field in these ranges has a significant impact on their performance.
Based on these insights, we configured network classifiers that achieved a number of top results in several DCASE 2019 challenge tasks~\cite{Koutinitrrfcnns2019}, as will be briefly reported in Section \ref{sec:dcase2019}.

\section{Regularizing CNN Architectures and Introducing Frequency Awareness}
\label{sec:arcs}

%!TEX root = ../main.tex

\begin{table}[t]
\caption{Modified ResNet architectures}
\begin{center}
\begin{tabular}{|c|c|}
\hline
\textbf{RB Number}&\textbf{RB Config} \\
%\cline{2-3} 
%\textbf{} & \textbf{\textit{RN1}}& \textbf{\textit{RN2}} \\
\hline
&Input $ 5 \times 5$ stride=$2$ 
\\
\hline

1&$3 \times 3$, $ 1 \times 1$, P\\
2  & $ x_1 \times x_1$,  $ x_2 \times x_2$, P  \\
3  & $ x_3 \times x_3$,  $ x_4 \times x_4$  \\
4    & $ x_5 \times x_5$,  $ x_6 \times x_6$, P   \\
5&$x_7 \times x_7$, $ x_8 \times x_8$  \\
6 &$ x_9 \times x_9$, $ x_{10} \times x_{10}$  \\
7 &$ x_{11} \times x_{11}$, $ x_{12} \times x_{12}$ \\
8  &$ x_{13} \times x_{13}$, $ x_{14} \times x_{14}$  \\
9 &$ x_{15} \times x_{15}$, $ x_{16} \times x_{16}$  \\
10&$ x_{17} \times x_{17}$, $ x_{18} \times x_{18}$  \\
11 &$ x_{19} \times x_{19}$, $ x_{20} \times x_{20}$  \\
12  &$ x_{21} \times x_{21}$, $ x_{22} \times x_{22}$  \\
\hline
\multicolumn{2}{l}{RB: Residual Block, P: $ 2 \times 2$ max pooling after the block.}\\
\multicolumn{2}{l}{$x_k \in \{ 1 , 3 \}$: hyper parameter we use to control the RF}\\
\multicolumn{2}{l}{ of the network. Number of channelds per RB:}\\
\multicolumn{2}{l}{128 for RBs 1-4; 256 for RBs 5-8; 512 for RBs 9-12.} % \\
% \multicolumn{2}{l}{RB number 5-8 have 256 channels.}\\
% \multicolumn{2}{l}{RB number 9-12 have 512 channels.}
\end{tabular}
\label{tab_resnet_configs}
\end{center}
\end{table}

\begin{table}[t]
\caption{Mapping $\rho$ values to the maximum RF of ResNet variants
(networks configured as in Table~\ref{tab_resnet_configs}).
$\rho$ controls the maximum RF by setting the $x_k$ as explained in Eq.~\eqref{eqn:rfset}.}
\begin{center}
\begin{tabular}{|c|c||c|c|}
\hline
\textbf{ $\rho$ value}&\textbf{Max RF}&\textbf{ $\rho$ value}&\textbf{Max RF} \\ \hline
%\cline{2-3} 
%\textbf{} & \textbf{\textit{RN1}}& \textbf{\textit{RN2}} \\
0 & $ 23  \times  23  $
&
1 & $ 31  \times  31  $
\\ \hline
2 & $ 39  \times  39  $
&
3 & $ 55  \times  55  $
\\ \hline
4 & $ 71  \times  71  $
&
5 & $ 87  \times  87  $
\\ \hline
6 & $ 103  \times  103  $
&
7 & $ 135  \times  135  $
\\ \hline
8 & $ 167  \times  167  $
&
9 & $ 199  \times  199  $
\\ \hline
10 & $ 231  \times  231  $
&
11 & $ 263  \times  263  $
\\ \hline
12 & $ 295  \times  295  $
&
13 & $ 327  \times  327  $
\\ \hline
14 & $ 359  \times  359  $
&
15 & $ 391  \times  391  $
\\ \hline
16 & $ 423  \times  423  $
&
17 & $ 455  \times  455  $
\\ \hline
18 & $ 487  \times  487  $
&
19 & $ 519  \times  519  $
\\ \hline
20 & $ 551  \times  551  $
&
21 & $ 583  \times  583  $
\\ \hline
\hline
% \multicolumn{4}{l}{$\rho$ controls the maximum RF of the network}\\
% \multicolumn{4}{l}{The network in configured as in Table~\ref{tab_resnet_configs},}\\
% \multicolumn{4}{l}{$\rho$ controls $x_k$ as explained in~\eqref{eqn:rfset}.}\\
\end{tabular}
\label{tab_rho_phi}
\end{center}
\end{table}

As shown in our previous work~\cite{Koutini2019Receptive}, the size of the receptive field (RF) is crucial when applying CNNs to audio recognition tasks. Following the
guidelines in~\cite{Koutini2019Receptive}, we adapted various ResNet and DenseNet variants. % following the guidelines of our previous work~\cite{Koutini2019Receptive}.
Using the provided development set for Task 1.A~\cite{MesarosDCASE2019T1}, we performed a grid search on the RF of the various ResNet architectures and
% We selected the following ResNet variants in computer vision (Section~\ref{sec:intro}).
show the performance of different CNNs under different RF setups. The goal of this investigation, reported in Section \ref{sec:resnet}, is to introduce a method to systematically push the performance of a single CNN architecture for acoustic scene classification. We base on this method our submissions~\cite{Koutinitrrfcnns2019} to the DCASE 2019 challenge\cite{dcase2019web}, especially our top performing single architecture submission for Task 1.A (\emph{cp\_resnet}).

Furthermore, in Section \ref{sec:facnn} we introduce \emph{Frequency-aware CNNs} to address the possible shortcomings of models with a smaller receptive field. Systematic experiments will then show whether, or in what cases, this actually helps improve the results.
% We concluded that the optimal RF for the Task 1.A dataset is around $90\times90$ pixels of our extracted spectrograms (Section~\ref{sec:setup:dataprep}). 
% Figure~\ref{fig:resultsrf} shows the validation loss of the provided development set, for ResNet with different RFs using the mono input.
% The steps taken to produce this search will be published at \url{https://github.com/kkoutini/cpjku_dcase19}.
 %(maybe a plot would be nice here or maybe in a follow up workhsop paper)

% \begin{figure}[h]
%   \centering
%   \centerline{\includegraphics[width=\columnwidth]{valloss.png}}
%   \caption{Validation Loss of the provided development split of Task 1 a dataset, for ResNet with different receptive fields over a mono input.}
%   \label{fig:resultsrf}
% \end{figure}

% In Task 1.B, we used network architectures with the same receptive field as in Task 1.A, motivated by the similarity and overlap between the two datasets, as they are both about modeling the acoustic scenes.
% 
% For Task 2, we searched for the optimal RF using a 4 fold Cross-Validation (CV) of the curated data only. We found that the optimal receptive field for the dataset is around $100\times100$ pixels over the extracted spectrograms.
\subsection{Adapting the Receptive Field of CNNs}

\subsubsection{ResNet}
\label{sec:resnet}
ResNet~\cite{heDeepResidualLearning2016} and its variants (such as preact-ResNet~\cite{He2016preact}) achieve state-of-the-art results in image recognition.
As we show in our recent study~\cite{Koutini2019Receptive}, such architectures can be adapted to audio tasks using RF regularization.
We adapt the RF of the ResNet in a similar fashion to ~\cite{Koutini2019Receptive} as explained below. The resulting network architectures are detailed in Table~\ref{tab_resnet_configs}. We use the hyper-parameters $x_k \in \{ 1 , 3 \}$, corresponding to filter sizes at different CNN levels (see Fig.~\ref{tab_resnet_configs}), to control the RF of the network. In order to simplify the process of adjusting the RF of the network, we introduce a new hyper-parameter $\rho$. We use $\rho$ to control $x_k$ as explained in ~\eqref{eqn:rfset}.
\begin{equation}
  \label{eqn:rfset}
 x_k = 
     \begin{cases}
        3 &\quad\text{if }k\le \rho \\
       1 &\quad\text{if } k > \rho \\
     \end{cases}
\end{equation}
For example, setting $\rho=5$ will result in a ResNet configured as in Table~\ref{tab_resnet_configs} with $x_k=3 \text{ for }k \in [1,5]$ and  $x_k=1 \text{ otherwise}$. The resulting ResNet has a RF of $87\times87$. Table~\ref{tab_rho_phi} maps $\rho$ values to the maximum RF of the resulting network\footnote{We will release the source code used to produce these networks and replicate the experiments at \url{https://github.com/kkoutini/cpjku_dcase19}}.

Networks with larger receptive fields degrade in performance as shown in~\cite{Koutini2019Receptive}. For this reason, we present the results of $\rho$ values in the range $\rho \in [1,12]$
% Resnet consist of 
\subsubsection{PreAct ResNet}
\label{sec:preact}
PreAct ResNet is a ResNet variant where residual branches are summed up before applying the non-linearity~\cite{He2016preact}. We specifically use \textit{PreActBN} as explained in~\cite{preactbn2018}, since it improves the performance of vanilla PreAct ResNet with and without Shake-Shake regularization for speech emotion recognition.

We control the RF of PreAct ResNets in the same manner as ResNets (Section~\ref{sec:resnet}). Table~\ref{tab_resnet_configs} and Equation~\ref{eqn:rfset} explain the configurations of our tested networks.

 \begin{figure*}[t]
  \centering
  \centerline{\includegraphics[width=\textwidth]{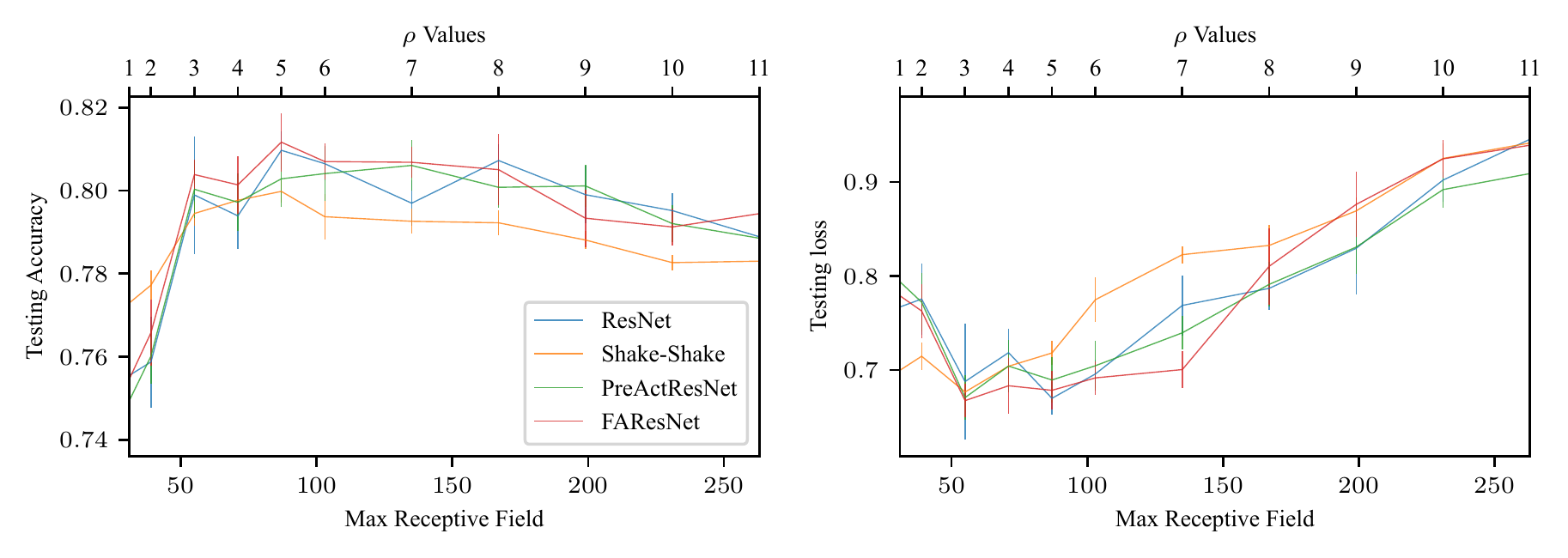}}
  \caption{Testing Loss/Accuracy of the provided development split of Task 1 a dataset, for  ResNet variants with different receptive fields over the input \textbf{without} mix-up.}
  \label{fig:results_nomixup}
\end{figure*}

\subsubsection{Shake-Shake ResNet}
\label{sec:shake}

The Shake-Shake architecture~\cite{gastaldi2017shake} is a variant of ResNet that is proposed for improved stability and robustness.
Each residual block has 3 branches; an identity map of the input and 2 convolutional branches, which are summed with random coefficients (in both the forward and backward pass)~\cite{gastaldi2017shake}. This regularization technique has shown empirically to improve the performance of CNNs on many tasks. We also specifically use Shake-Shake regularized \textit{PreActBN} from~\cite{preactbn2018}. In Shake-Shake regularized ResNets, each residual block only has a new branch that is added to the sum. Therefore, the resulting maximum RF of the network is not changed. In result, we use the same techniques to control the RF (Section~\ref{sec:resnet}). Table~\ref{tab_resnet_configs} shows the configuration of both branches of the residual blocks. 

Although, Shake-Shake ResNet is not performing well in the classic acoustic scene classification problem (as shown in Section~\ref{sec:results}), it excels in the case of domain mismatch (Task 1.B~\cite{MesarosDCASE2019T1,dcase2019web}) and noisy datasets (Task 2~\cite{fonseca2019audio})~\cite{Koutinitrrfcnns2019}.
\subsubsection{DenseNet}
\label{sec_dn}

We adapted DenseNet~\cite{huangDenselyConnectedConvolutional2017} in a similar fashion to DN1 in ~\cite{Koutini2019Receptive}. We report on two DenseNet configurations with maximum RF of $87 \times 87$ and $71 \times 71$ pixels (Section~\ref{sec:results}).
 % solves the vanishing gradient problem by concatenating the outputs of all previous convolution layers on the channels dimension. DenseNet variants achieves state-of-the-art results on computer vision tasks.

\subsection{Frequency-aware Convolution}
\label{sec:facnn}
In CNNs that have a large enough RF, deeper convolutional layers can infer the frequency information of their input feature maps. However, CNNs with large RF degrade in performance and fail to generalize in acoustic scene classification as shown in~\cite{Koutini2019Receptive}.
On the other hand, in high-performing fully convolutional CNNs, learned CNN filters are agnostic to the frequency range information of the feature maps. In other words, the spectrograms and feature maps can be rolled over both the time and frequency dimension with a minor impact on the network predictions. This is one side effect of limiting the receptive field of CNNs on spectograms. %is that convolutional layers have less inferred information about the frequency context of the feature maps. In addition to that, 
We propose a new convolutional layer, which we call \emph{Frequency-aware Convolution}, to make filters aware and more specialized in certain frequencies by concatenating a new channel containing the frequency information of each spatial pixel to each feature map. This technique is similar to CoordConv~\cite{liu2018intriguing}, where the network input is padded with the pixels' coordinates. In our case, we pad all feature maps with a real number indicating the frequency context of each pixel.\footnote{In this paper, we used a number between $-1$ and $1$, where $-1$ represents the lowest and $1$ the highest frequency in the spectrogram. But this range can be adapted according to the value range of the input.}

The CNN models that incorporate our frequency-aware layer will be called the \emph{Frequency-Aware Convolutional Neural Networks (FACNNs)}. Similarly, we call the frequency-aware ResNet \emph{FAResNet}.
We denote the value of the pixel with spatial index $(f,t)$ in the new channel as $V(f,t)$; it is calculated as
\begin{equation}
  \label{eqn:locpad}
    V(f,t)=f/F
\end{equation}
where $F$ is the size of the frequency dimension of the feature map, $f$ is the pixel index in the frequency dimension, and $t$ is the pixel index in the time dimension.
This new channel gives the convolutional filters a frequency context.

Since making CNNs frequency-aware (by adding the new channel) does not alter the maximum RF of the network, we control the maximum RF of FAResNets similar to ResNet (Section~\ref{sec:resnet}) by changing $\rho$.

\section{Experimental Setup}
\label{sec:setup}

\begin{figure*}[t]
  \centering
  \centerline{\includegraphics[width=\textwidth]{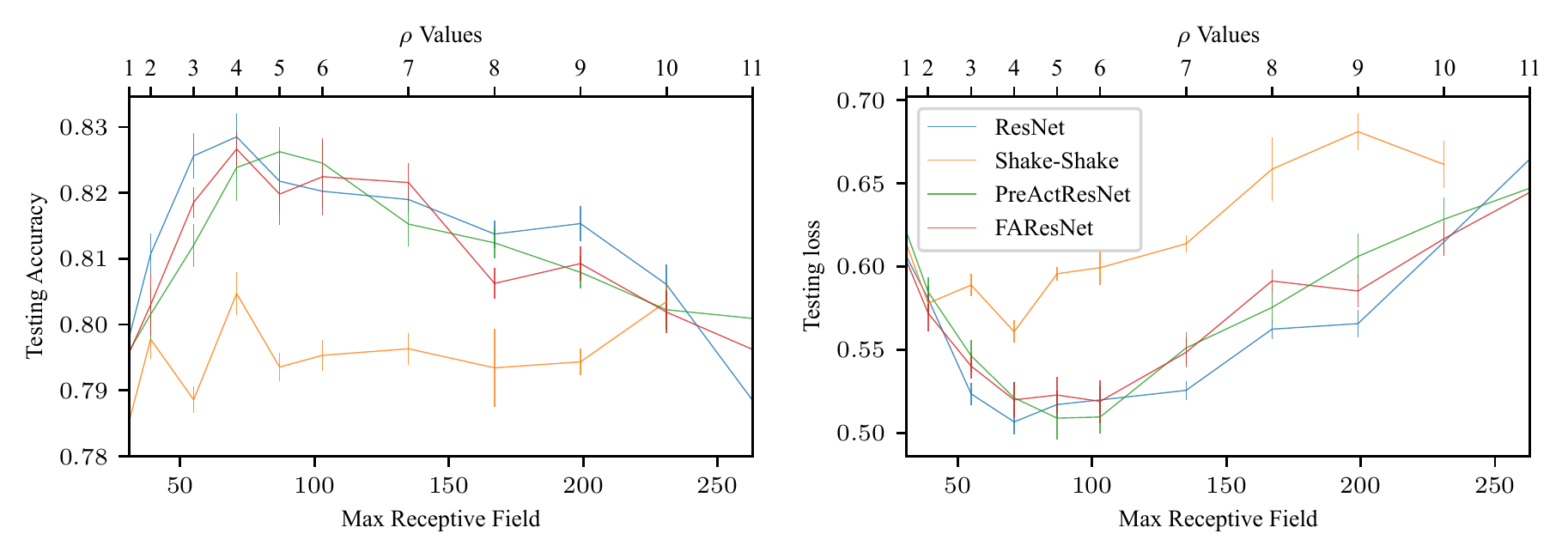}}
  \caption{Testing Loss/Accuracy of the provided development split of Task 1 a dataset, for  ResNet variants with different receptive fields over the input \textbf{with} mix-up.}
  \label{fig:results_mixup}
  \vspace{-4mm}
\end{figure*}

%!TEX root = ../main.tex

% TODO rewrite

\subsection{Data Preparation and Training}
\label{sec:setup:dataprep}
%The input is and subjected to a 
We extracted the input features using a Short Time Fourier
Transform (STFT) with a window size of 2048 and 25\% overlap. We perceptually weight the resulting spectrograms and apply a Mel-scaled filter bank in a similar fashion to Dorfer et al.~\cite{DorferDCASE2018task1}. This preprocessing results in 256 Mel frequency bins. The input is first down-sampled to 22.05 kHz. 
We process each input channel of the stereo audio input independently and provide the CNN with a two-channel-spectrogram input.
The input frames are normalized using the training set mean and standard deviation.

% The input frames are normalized %using the training set mean and standard deviation.
% to zero-mean and unit variance according to the training set.

We used Adam~\cite{kingmaAdamMethodStochastic2014}  with a specific scheduler. We start training with a learning rate of $1 \times 10^{-4}$. From epoch 50 until 250, the learning rate decays linearly from $1 \times 10^{-4}$ to  $5 \times 10^{-6}$. We train for another 100 epochs with the minimum learning rate $5 \times 10^{-6}$ in a setup similar to~\cite{Koutini2019Receptive}.

\subsection{Testing Setup}
We use the provided development split of DCASE 2019 task 1A~\cite{MesarosDCASE2019T1,dcase2019web}. We train our models on the provided training set and treat the provided test set as unseen set. In other words, we don't select best models based on their performance  on the test set. Instead, for each model, we report the average results of the last 25 training epochs of two runs on the test set.

\subsection{Data Augmentation}
\textbf{Mix-Up:}  \textit{Mix-up}~\cite{zhangMixupEmpiricalRisk2017} is an effective augmentation method that works by linearly combining two input samples and their targets. It was shown to have a great impact on the performance and the generalization of the models.\newline  \textbf{Spectogram Rolling:} We roll the spectograms randomly over the time dimension.

% \subsection{Model Averaging}
% \textbf{Stochastic Weight Averaging:} Stochastic Weight Averaging (SWA)~\cite{izmailov2018averaging} had led to a better performance on the validation set in our experiments on Task 2. We keep an SWA copy of the model parameters while training.
% This SWA average is updated with new parameters every 3 epochs.
% \\\textbf{Snapshot Averaging}: A snapshot of the model during training is saved every 5 epochs, within the last 100 epochs.
% The predictions of all these models are then averaged for the final prediction.
% The snapshot averaging is computationally more efficient than training separate models, and easier than stochastic weight averaging approaches such as~\cite{izmailov2018averaging} for creating ensemble models, as no re-computation of the layer statistics (such as batch-norm) is required.
% \\\textbf{CV Averaging}: A model is trained on different CV folds, and the predictions of models from different folds are averaged. This approach provides more robustness, as different models have seen different training data points, hence found slightly different minima. This approach is simple yet effective, and was used successfully in our previous submissions~\cite{eghbal-zadehCPJKUSubmissionsDCASE20162016,DorferDCASE2018task1,lehnerClassifyingShortAcoustic2017}.

\section{Results}
\label{sec:results}

 Table~\ref{tab_results} shows the $\rho$ and Max RF configuration that achieves the top accuracy (mean/std over the last 25 epochs) for each architecture, with and without mix-up. It is also worth noting that the \emph{maximum} RF is different from the \emph{effective} RF as explained in ~\cite{Koutini2019Receptive,luoUnderstandingEffectiveReceptive2016}. We control the maximum RF using $\rho$, while the \emph{effective} RF is dependent on the architecture, the initialization and the data~\cite{luoUnderstandingEffectiveReceptive2016}. 
This is one possible explanation for why different architectures may have a slightly shifted optimal maximum RF range (for example, PreAct in Table~\ref{tab_results} and Figure~\ref{fig:results_mixup}). Likewise, using mix-up can alter the optimal maximum RF range for the networks.

\subsection{Without Mix-up}
Figure~\ref{fig:results_nomixup} shows the testing loss and accuracy for different architectures % (Section~\ref{sec:arcs})
over a range of $\rho$ values and -- consequently (see Eq.~\eqref{eqn:rfset}) -- maximum RF values. The plots summarize the results for the last 25 epochs of 2 runs. We notice that FAResNet excels mostly in smaller RF networks ($\rho < 8$) where frequency context is more valuable. The figure also shows the best-performing maximum RF range for different architectures to correspond to $\rho$ values in the range $[3,8]$. In this range, FAResNet outperforms other ResNet variants.

\subsection{With  Mix-up}
Figure~\ref{fig:results_mixup} shows the testing loss and accuracy
% similarly to Figure~\ref{fig:results_nomixup}
when we use mix-up data augmentation. We note that when using mix-up, ResNet outperforms the other variants. Further experiments and investigation are still needed to fully understand the effect of mix-up on these architectures. The figure shows that the best-performing maximum RF range for architectures corresponds to $\rho$ values in the range $[3,5]$ for \emph{ResNet} and \emph{FAResNet}, and $[4,6]$  for  \emph{PreActResnet}. Shake-Shake achieves its best performance for $\rho=4$. We see that performance degrades outside these maximum RF ranges for different architectures, in accordance with~\cite{Koutini2019Receptive}.

%\gerhard{We need a bit more discussion/analysis of the results here ...}

\begin{table} % [h]
\caption{Configurations with top accuracy per network architecture and its corresponding $\rho$ and max RF values with/without mix-up}
\begin{center}
\begin{tabular}{|c|c|c|c|c|}
\hline
\textbf{Network }& \textbf{$\rho$ }&\textbf{Max RF}&\textbf{ M/U}&\textbf{Accuracy} \\ \hline
%\cline{2-3} 
%\textbf{} & \textbf{\textit{RN1}}& \textbf{\textit{RN2}} \\
ResNet & 4 & $ 71  \times  71  $ & \cmark & $\textbf{82.85\%} \pm \textbf{.36} $\\ \hline
PreAct & 5 & $ 87  \times  87  $ & \cmark & $82.62\% \pm .37 $\\ \hline
Shake-Shake & 4 & $ 71  \times  71  $ & \cmark & $80.47\% \pm .32 $\\ \hline
FAResNet & 4 & $ 71  \times  71  $ & \cmark & $82.66\% \pm .27 $\\ \hline
DenseNet &  & $ 71  \times  71  $ & \cmark &   $ 81.53 \%   \pm .26 $ \\ \hline

ResNet & 5 & $ 87  \times  87  $ & \xmark & $80.97\% \pm .46 $\\ \hline
PreAct & 7 & $ 135  \times  135  $ & \xmark & $80.6\% \pm .61 $\\ \hline
Shake-Shake & 5 & $ 87  \times  87  $ & \xmark & $79.98\% \pm .27 $\\ \hline
FAResNet & 5 & $ 87  \times  87  $ & \xmark &   $ \textbf{ 81.17 \% }  \pm \textbf{.7} $ \\ \hline
DenseNet &  & $ 87  \times  87  $ & \xmark &   $ 79.9 \%   \pm .3 $ \\ \hline
\multicolumn{5}{l}{M/U: using Mix-Up}\\
\end{tabular}
\label{tab_results}
\end{center}
 \vspace{-8mm}
\end{table}

\subsection{Performance at DCASE 2019}
\label{sec:dcase2019}
Our receptive field regularized networks achieved the second place~\cite{Koutinitrrfcnns2019} (team ranking) in Task 1.A of the DCASE 2019 challenge~\cite{dcase2019web,MesarosDCASE2019T1}. We averaged ResNet, PreAct and FAResNet configured with $\rho=5$ to achieve $83.8 \%$ accuracy on the evaluation set. Our ResNet configured with $\rho=5$ (our single architecture submission \textit{cp\_resnet}) achieved $82.8 \%$ accuracy when trained on the whole development set; we averaged the prediction of the last training epochs~\cite{Koutinitrrfcnns2019}. When instead averaging the predictions of the same architecture trained on a 4-fold cross-validation of the development data, it achieves $83.7 \%$ accuracy on the evaluation set. Furthermore, the submission achieved the highest accuracy on the unseen cities in the evaluation set ($78.1 \%$). 

The generality and robustness of the proposed RF regularization strategy is demonstrated by the fact that our highly-performing submissions to DCASE 2019 Tasks 1.B and 2~\cite{Koutinitrrfcnns2019} are also based on these architectures.

%\gerhard{This model (which?) was submitted to Task 1.a and achieved ...Also, models based on .. .. other tasks ... }

% \subsection{Ensembles}
% \khaled{(maybe?)}
% \hamid{if we are gonna report other tasks, this would be the place to report them all along with ensemble results from the challenge, and mention it ranked 2nd place in task1a, 3rd in task1b, 5th in task2, etc. in that case, we can rename this subsection to DCASE-2019 results.}

\section{Conclusion}
\label{sec:conc}
%!TEX root = ../main.tex
In this paper, we have investigated different configurations of deep CNN architectures that correspond to different maximum receptive fields over audio spectograms. We showed that this helps to better design deep CNNs for acoustic classification tasks, and to adapt CNNs performing well in other domains (notably, image recognition) to acoustic scene classification.
% In other words, we demonstrate the importance of the receptive field in regularizing deeps CNNs when working with spectograms. 
% We report the results on DCASE 2019 task 1.a dataset. In addition, this paper explains the experiments we performed to produce our top performing submissions to DCASE 2019 task 1.
The good results achieved with this basic strategy in several DCASE 2019 tasks suggest that this is a very general and robust approach that may prove beneficial in
% architectures can be an effective tool that offer good generalization properties for
various other audio processing tasks.

%Classwise RF 

%Ensembele of different nets with different RFs

\section{ACKNOWLEDGMENT}
\label{sec:ack}

This work has been supported by the COMET-K2 Center of the Linz Center of Mechatronics (LCM) funded by the Austrian federal government and the Federal State of Upper Austria.

% -------------------------------------------------------------------------
% Either list references using the bibliography style file IEEEtran.bst
\bibliographystyle{IEEEtran}
\bibliography{main}

\begin{thebibliography}{10}
\providecommand{\url}[1]{#1}
\def\UrlFont{\rmfamily}
\providecommand{\newblock}{\relax}
\providecommand{\bibinfo}[2]{#2}
\providecommand\BIBentrySTDinterwordspacing{\spaceskip=0pt\relax}
\providecommand\BIBentryALTinterwordstretchfactor{4}
\providecommand\BIBentryALTinterwordspacing{\spaceskip=\fontdimen2\font plus
\BIBentryALTinterwordstretchfactor\fontdimen3\font minus
  \fontdimen4\font\relax}
\providecommand\BIBforeignlanguage[2]{{%
\expandafter\ifx\csname l@#1\endcsname\relax
\typeout{** WARNING: IEEEtran.bst: No hyphenation pattern has been}%
\typeout{** loaded for the language `#1'. Using the pattern for}%
\typeout{** the default language instead.}%
\else
\language=\csname l@#1\endcsname
\fi
#2}}

\bibitem{Koutini2019Receptive}
K.~Koutini, H.~Eghbal-zadeh, M.~Dorfer, and G.~Widmer, ``{The Receptive Field
  as a Regularizer in Deep Convolutional Neural Networks for Acoustic Scene
  Classification},'' in \emph{Proceedings of the European Signal Processing
  Conference (EUSIPCO)}, A Coru\~{n}a, Spain, 2019.

\bibitem{eghbal-zadehCPJKUSubmissionsDCASE20162016}
H.~Eghbal-Zadeh, B.~Lehner, M.~Dorfer, and G.~Widmer, ``{{CP}}-{{JKU}}
  submissions for {{DCASE}}-2016: {{A}} hybrid approach using binaural
  i-vectors and deep convolutional neural networks,'' in \emph{{{DCASE}}
  2016-{{challenge}} on {{Detection}} and {{Classification}} of {{Acoustic
  Scenes}} and {{Events}}}.\hskip 1em plus 0.5em minus 0.4em\relax {DCASE2016
  Challenge}, 2016.

\bibitem{hersheyCNNArchitecturesLargescale2017}
S.~Hershey, S.~Chaudhuri, D.~P.~W. Ellis, J.~F. Gemmeke, A.~Jansen, R.~C.
  Moore, M.~Plakal, D.~Platt, R.~A. Saurous, B.~Seybold, M.~Slaney, R.~J.
  Weiss, and K.~Wilson, ``{{CNN}} architectures for large-scale audio
  classification,'' in \emph{2017 {{IEEE International Conference}} on
  {{Acoustics}}, {{Speech}} and {{Signal Processing}} ({{ICASSP}})}, 2017, pp.
  131--135.

\bibitem{lehnerClassifyingShortAcoustic2017}
B.~Lehner, H.~Eghbal-Zadeh, M.~Dorfer, F.~Korzeniowski, K.~Koutini, and
  G.~Widmer, ``Classifying short acoustic scenes with {{I}}-vectors and
  {{CNNs}}: {{Challenges}} and optimisations for the 2017 {{DCASE ASC}} task,''
  in \emph{{{DCASE}} 2017-{{challenge}} on {{Detection}} and {{Classification}}
  of {{Acoustic Scenes}} and {{Events}}}.\hskip 1em plus 0.5em minus
  0.4em\relax {DCASE2017 Challenge}, 2017.

\bibitem{DorferDCASE2018task1}
M.~Dorfer, B.~Lehner, H.~Eghbal-zadeh, C.~Heindl, F.~Paischer, and G.~Widmer,
  ``Acoustic {{scene classification}} with {{fully convolutional neural
  networks}} and {{I}}-{{vectors}},'' in \emph{Proceedings of the {{Detection}}
  and {{Classification}} of {{Acoustic Scenes}} and {{Events}} 2018
  {{Challenge}} ({{DCASE2018}})}, 2018.

\bibitem{SakashitaDCASE2018task1}
Y.~Sakashita and M.~Aono, ``Acoustic {{scene classification}} by {{ensemble}}
  of {{spectrograms based}} on {{adaptive temporal divisions}}.''\hskip 1em
  plus 0.5em minus 0.4em\relax {DCASE2018 Challenge}, 2018.

\bibitem{DorferDCASE2018task2}
M.~Dorfer and G.~Widmer, ``Training general-purpose audio tagging networks with
  noisy labels and iterative self-verification,'' in \emph{Proceedings of the
  {{Detection}} and {{Classification}} of {{Acoustic Scenes}} and {{Events}}
  2018 {{Workshop}} ({{DCASE2018}})}, 2018, pp. 178--182.

\bibitem{IqbalDCASE2018task2}
T.~Iqbal, Q.~Kong, M.~Plumbley, and W.~Wang, ``Stacked {{convolutional neural
  networks}} for {{general}}-purpose {{audio tagging}}.''\hskip 1em plus 0.5em
  minus 0.4em\relax {DCASE2018 Challenge}.

\bibitem{LeeDCASE2017task4}
D.~Lee, S.~Lee, Y.~Han, and K.~Lee, ``Ensemble of convolutional neural networks
  for weakly-supervised sound event detection using multiple scale
  input.''\hskip 1em plus 0.5em minus 0.4em\relax {DCASE2017 Challenge}.

\bibitem{KoutiniDCASE2018task4}
K.~Koutini, H.~Eghbal-zadeh, and G.~Widmer, ``Iterative knowledge distillation
  in {{R}}-{{CNNs}} for weakly-labeled semi-supervised sound event detection,''
  in \emph{Proceedings of the {{Detection}} and {{Classification}} of
  {{Acoustic Scenes}} and {{Events}} 2018 {{Workshop}} ({{DCASE2018}})}, 2018,
  pp. 173--177.

\bibitem{heDeepResidualLearning2016}
K.~He, X.~Zhang, S.~Ren, and J.~Sun, ``Deep residual learning for image
  recognition,'' in \emph{Proceedings of the {{IEEE}} Conference on Computer
  Vision and Pattern Recognition}, 2016, pp. 770--778.

\bibitem{huangDenselyConnectedConvolutional2017}
G.~Huang, Z.~Liu, L.~Van Der~Maaten, and K.~Q. Weinberger, ``Densely connected
  convolutional networks,'' in \emph{Proceedings of the {{IEEE}} Conference on
  Computer Vision and Pattern Recognition}, 2017, pp. 4700--4708.

\bibitem{Koutinitrrfcnns2019}
K.~Koutini, H.~Eghbal-zadeh, and G.~Widmer, ``Acoustic scene classification and
  audio tagging with receptive-field-regularized {CNNs},'' DCASE2019 Challenge,
  Tech. Rep., June 2019.

\bibitem{brendel2018approximating}
\BIBentryALTinterwordspacing
W.~Brendel and M.~Bethge, ``Approximating {CNN}s with bag-of-local-features
  models works surprisingly well on {{ImageNet}},'' in \emph{International
  Conference on Learning Representations}, 2019. [Online]. Available:
  \url{https://openreview.net/forum?id=SkfMWhAqYQ}
\BIBentrySTDinterwordspacing

\bibitem{sabour2017dynamic}
S.~Sabour, N.~Frosst, and G.~E. Hinton, ``Dynamic routing between capsules,''
  in \emph{Advances in neural information processing systems}, 2017, pp.
  3856--3866.

\bibitem{liu2018intriguing}
R.~Liu, J.~Lehman, P.~Molino, F.~P. Such, E.~Frank, A.~Sergeev, and
  J.~Yosinski, ``An intriguing failing of convolutional neural networks and the
  coordconv solution,'' in \emph{Advances in Neural Information Processing
  Systems}, 2018, pp. 9605--9616.

\bibitem{He2016preact}
K.~He, X.~Zhang, S.~Ren, and J.~Sun, ``Identity mappings in deep residual
  networks,'' \emph{arXiv preprint arXiv:1603.05027}, 2016.

\bibitem{preactbn2018}
\BIBentryALTinterwordspacing
C.~Huang and S.~S. Narayanan, ``Normalization before shaking toward learning
  symmetrically distributed representation without margin in speech emotion
  recognition,'' \emph{CoRR}, vol. abs/1808.00876, 2018. [Online]. Available:
  \url{http://arxiv.org/abs/1808.00876}
\BIBentrySTDinterwordspacing

\bibitem{gastaldi2017shake}
X.~Gastaldi, ``Shake-shake regularization,'' \emph{arXiv preprint
  arXiv:1705.07485}, 2017.

\bibitem{MesarosDCASE2019T1}
A.~Mesaros, T.~Heittola, and T.~Virtanen, ``Acoustic scene classification in
  {{DCASE}} 2019 challenge: Closed and open set classification and data
  mismatch setups,'' in \emph{Proceedings of the {{Detection}} and
  {{Classification}} of {{Acoustic Scenes}} and {{Events}} 2019 {{Workshop}}
  ({{DCASE2019}})}, 2019.

\bibitem{dcase2019web}
\url{http://dcase.community/challenge2019/task-acoustic-scene-classification}.

\bibitem{fonseca2019audio}
E.~Fonseca, M.~Plakal, F.~Font, D.~P. Ellis, and X.~Serra, ``Audio tagging with
  noisy labels and minimal supervision,'' \emph{arXiv preprint
  arXiv:1906.02975}, 2019.

\bibitem{kingmaAdamMethodStochastic2014}
D.~P. Kingma and J.~Ba, ``Adam: {A} method for stochastic optimization,'' in
  \emph{3rd International Conference on Learning Representations, {ICLR} 2015,
  San Diego, CA, USA, May 7-9, 2015, Conference Track Proceedings}, 2015.

\bibitem{zhangMixupEmpiricalRisk2017}
H.~Zhang, M.~Ciss{\'{e}}, Y.~N. Dauphin, and D.~Lopez{-}Paz, ``mixup: Beyond
  empirical risk minimization,'' in \emph{6th International Conference on
  Learning Representations, {ICLR} 2018, Vancouver, BC, Canada, April 30 - May
  3, 2018, Conference Track Proceedings}, 2018.

\bibitem{luoUnderstandingEffectiveReceptive2016}
W.~Luo, Y.~Li, R.~Urtasun, and R.~Zemel, ``Understanding the {{Effective
  Receptive Field}} in {{Deep Convolutional Neural Networks}},'' in
  \emph{Advances in {{Neural Information Processing Systems}} 29}, 2016, pp.
  4898--4906.

\end{thebibliography}
%
% or list them by yourself
% \begin{thebibliography}{9}
% 
% \bibitem{dcase2016web}
%   \url{http://www.cs.tut.fi/sgn/arg/dcase2016/}.
%
% \bibitem{IEEEPDFSpec}
%   {PDF} specification for {IEEE} {X}plore$^{\textregistered}$,
%   \url{http://www.ieee.org/portal/cms_docs/pubs/confstandards/pdfs/IEEE-PDF-SpecV401.pdf}.
%
% \bibitem{PDFOpenSourceTools}
%   Creating high resolution {PDF} files for book production with 
%   open source tools, 
%   \url{http://www.grassbook.org/neteler/highres_pdf.html}.
%
% \bibitem{eWilliams1999}
% E. Williams, \emph{Fourier Acoustics: Sound Radiation and Nearfield Acoustic
%   Holography}. London, UK: Academic Press, 1999.
% 
% \bibitem{ieeecopyright}
%   \url{http://www.ieee.org/web/publications/rights/copyrightmain.html}.
%
% \bibitem{cJones2003}
% C. Jones, A. Smith, and E. Roberts, ``A sample paper in conference
%   proceedings,'' in \emph{Proc. IEEE ICASSP}, vol. II, 2003, pp. 803--806.
% 
% \bibitem{aSmith2000}
% A. Smith, C. Jones, and E. Roberts, ``A sample paper in journals,'' 
%   \emph{IEEE Trans. Signal Process.}, vol. 62, pp. 291--294, Jan. 2000.
% 
% \end{thebibliography}

\end{sloppy}
\end{document}